\def\Mdot {\mathaccent 95 M}

\def\Msun {\rm M_{\odot}}
\def\peryr{\rm yr^{-1}}
\def\B*   {B_{\rm *}}
\def\R*   {R_{\rm *}}
\def\M*   {M_{\rm *}}
\def\m*   {m_{\rm *}}

\def\O*   {\Omega _{\rm *}}

\def\NH   {N_{\rm H}}
\newbox\grsign \setbox\grsign=\hbox{$>$} 
\newdimen\grdimen \grdimen=\ht\grsign
\newbox\laxbox \newbox\gaxbox
\setbox\gaxbox=\hbox{\raise.5ex\hbox{$>$}\llap
     {\lower.5ex\hbox{$\sim$}}}\ht1=\grdimen\dp1=0pt
\setbox\laxbox=\hbox{\raise.5ex\hbox{$<$}\llap
     {\lower.5ex\hbox{$\sim$}}}\ht2=\grdimen\dp2=0pt
\def\gax{\mathrel{\copy\gaxbox}}

 \documentstyle[12pt,aasms4]{article}

\lefthead{Mauche}
\righthead{Oscillations in the EUV flux of SS~Cygni}

\begin{document}

\title{Quasi-Coherent Oscillations in the Extreme\\
       Ultraviolet Flux of the Dwarf Nova SS~Cygni}

\author{Christopher W.\ Mauche}
\affil{Lawrence Livermore National Laboratory,\\
       L-41, P.O.\ Box 808, Livermore, CA 94550;\\
       mauche@cygnus.llnl.gov}

\begin{abstract}
Quasi-coherent oscillations have been detected in the extreme ultraviolet
flux of the dwarf nova SS~Cygni during observations with the {\it Extreme
Ultraviolet Explorer\/} satellite of the rise and plateau phases of an
anomalous outburst in 1993 August and a normal outburst in 1994 June/July.
On both occasions, the oscillation turned on during the rise to outburst
and persisted throughout the observation. During the 1993 outburst, the
period of the oscillation fell from 9.3~s to 7.5~s over an interval of 4.4
days; during the 1994 outburst, the period fell from 8.9~s to 7.19~s (the
shortest period ever observed in SS~Cyg, or any other dwarf nova) within 
less than a day, and then rose to 8.0~s over an interval of 8.0 days. For
both outbursts, the period $P$ of the oscillation was observed to correlate
with the 75--120~\AA \ count rate $I_{\rm EUV}$ according to $P\propto
I_{\rm EUV}^{-0.094}$. A magnetospheric model is considered to reproduce
this variation. It is found that an effective high-order multipole field
is required, and that the field strength at the surface of the white dwarf
is 0.1--1~MG. Such a field strength is at the lower extreme of those measured
or inferred for bona~fide magnetic cataclysmic variables.
\end{abstract}


\clearpage   

\section{Introduction}

Rapid periodic oscillations are observed in the optical flux of high accretion
rate (``high-$\Mdot $'') cataclysmic variables (CVs; nova-like variables and
dwarf novae in outburst) (\cite{pat81}; \cite{war95}; \cite{war96}) These
oscillations have high coherence ($Q\approx 10^4$--$10^6$), periods $P\approx
10$--30~s, amplitudes of less than 0.5\%, and are sinusoidal to within the
limits of measurement. They are referred to as ``dwarf nova oscillations''
(DNOs) to distinguish them from the apparently distinct longer period, low
coherence ($Q\approx 1$--$10$) quasi-periodic oscillations (QPOs) of
high-$\Mdot$ CVs, and the longer period, high coherence ($Q\approx
10^{10}$--$10^{12}$) oscillations of DQ~Her stars. DNOs have never been detected
in dwarf novae in quiescence, despite extensive searches; they appear on the
rising branch of the dwarf nova outburst, typically persist through maximum,
and disappear on the declining branch of the outburst. The period of the
oscillation decreases on the rising branch and increases on the declining
branch, but because the period reaches a minimum about one day after maximum
optical flux, dwarf novae describe a loop in a plot of period versus optical
flux.

The dwarf nova SS~Cygni routinely exhibits DNOs during outburst. Optical
oscillations have been detected at various times with periods ranging from
8.2~s to 10.9~s (\cite{pat78}; \cite{hor80}; \cite{pat81}). During one outburst,
the period was observed over a interval of $\approx 6$ days to fall from
7.5~s to 7.3~s and then rise to 8.5~s (\cite{hil81}). At soft X-ray energies
($E\approx 0.1$--0.5 keV), oscillations have been detected in \hbox{{\it HEAO
1\/}} LED~1 data at periods of $\approx 9$~s and 11~s (\cite{cor80};
\cite{cor84}) and in {\it EXOSAT\/} LE data at periods between 7.4~s and 10.4~s
(\cite{jon92}). In this Letter, we describe observations with the {\it Extreme
Ultraviolet Explorer\/} satellite ({\it EUVE\/}; \cite{bow91}; \cite{bow94}) of
the EUV oscillations of SS~Cyg.

\section{Observations}

Target-of-opportunity observations of SS~Cyg in outburst were made with
{\it EUVE\/} in 1993 August (MJD 9216.58 to 9223.12; $\rm MJD =JD - 2440000$)
and 1994 June/July (MJD 9526.67 to 9529.78 and 9532.54 to 9536.94). On both
occasions, SS~Cyg was detected with the {\it EUVE\/} Deep Survey (DS)
photometer and Short Wavelength (SW) spectrometer. The optical and DS count
rate light curves of the outbursts are shown in Figure~1 of Mauche (1996a). On
both occasions, the optical flux was above $V=10$ for $\approx 16$ days, but
the 1993 outburst was anomalous in that it took $\approx 5$ days for the light
curve to reach maximum, whereas typical outbursts (such as the 1994 outburst)
reach maximum in 1 to 2 days. The EUV light curve of the 1993 outburst rose
more quickly than the optical, and the rise of the EUV light curve of the 1994
outburst was delayed by $\approx 1$ day relative to the optical. On both
occasions, the EUV spectrum evolved homologously over roughly two orders of
magnitude in luminosity. The EUV spectrum is very complex, but it can be very
crudely approximated by a blackbody with a temperature of $kT\approx 20$--30~eV
absorbed by a neutral hydrogen column density of $\NH\approx 7$--$4\times
10^{19}~\rm cm^{-2}$ (\cite{mau95}).

To detect oscillations in the EUV flux of SS~Cyg, we proceeded as follows.
First, we selected valid intervals when the source was above the Earth's limb
and both the DS and SW instruments were in operation. Only those intervals
longer than 600~s were retained for further consideration. For the 1993 (1994)
observations, 89 (75) valid intervals were defined with a net time of 155
(103) kiloseconds. Second, we calculated the background-subtracted count rate
in the DS and SW instruments for each of the valid time intervals. For the SW
instrument, only those photons between 75 and 120~\AA \ were included in
the sum; above and below these limits, the spectrum is dominated by noise
(\cite{mau95}). Third, for each of the valid time intervals, we constructed
background-subtracted DS count rate light curves with 1~s time resolution.
Fourth, we calculated the power spectra of these light curves using the XRONOS
software package. As with the {\it HEAO-1\/} (\cite{cor80}) and {\it EXOSAT\/}
(\cite{jon92}) soft X-ray power spectra, typical {\it EUVE\/} power spectra
consist of a single line in a single frequency bin, with no power above the
noise at any of the harmonics (\cite{mau6a}). From this result, we understand
that the EUV oscillation is (a) sinusoidal and (b) essentially coherent ($Q
\equiv |\Delta P/\Delta t|^{-1} > 6 \times 10^4$) over the $\sim 30$ min valid
time intervals. The period of the oscillation is then simply the inverse of the
frequency of the bin of the power spectrum with the peak power, and its error
is less than the bin width of $P^2/2048\sim 0.03$~s. Fifth, we calculated the
amplitude of the oscillation during each valid time interval by phase-folding
the data on the period appropriate to that interval and fitting a function of
the form $f(\phi )=A+B\sin (\phi + \phi_0)$ where $\phi = 2\pi t/P$. The
relative amplitude $B/A$ was $14\%\pm 12\%$ during the 1993 outburst and
$16\%\pm 12\%$ during the 1994 outburst, with a tendency in the 1994 data
for the relative amplitude to decrease as the count rate increased. These
amplitudes are comparable to those of the soft X-ray oscillations measured
by {\it HEAO-1\/} (\cite{cor80}) and {\it EXOSAT\/} (\cite{jon92}).

The period of the EUV oscillation of SS~Cyg detected by the above means
is plotted as a function of time in Figure~1. For the 1993 outburst, the
oscillation turned on during the rise to outburst on MJD 9218.77; its period
was initially 9.31~s, fell over the next $\approx 1.6$ days to $\approx 7.6$~s,
and then asymptotically approached $\approx 7.5$~s over the next few days. For
the 1994 outburst, the oscillation turned on during the fast rise to outburst
on MJD 9528.06; its period was initially 8.90~s, fell to 7.19~s (the shortest
period ever observed in SS~Cyg, or in any other dwarf nova) within less than a
day, and rebounded to 7.42~s by the end of the observation on MJD 9529.77. When
observations resumed 2.8 days later, the period of the oscillation was 7.59~s
and rose slowly over the next few days to $\approx 8.0$~s. On both occasions,
the oscillations switched on when the SW count rate was $\approx 0.1~\rm
counts~s^{-1}$ or when the DS count rate was $\sim 1~\rm count~s^{-1}$
(\cite{mau6a}). For a coherent oscillation to be detected below this count
rate, its amplitude must be $\gax 5\%$.

Superposed on Figure~1 is the log of the 75--120~\AA \ SW count rate as
a function of time. It is clear from this figure that the period of the
oscillation anti-correlates with the SW count rate, being long when the count
rate is low and short when the count rate is high. To quantify this trend,
we plot in Figure~2 the log of the period of the oscillation as a function
of the log of the SW count rate. The solid line connects the points
chronologically starting in the upper left-hand corner. The upper panel of
the figure shows the slow evolution of the period from 9.3~s to 7.5~s during
the 1993 outburst. The lower panel shows the initial rapid evolution of the
period during the 1994 outburst as the period fell from 8.9~s to 7.2~s and
then rose to 8.0~s. Two aspects of this plot require immediate comment.

First, despite the vastly different rate at which the period evolved in this
diagram during the 1993 and 1994 outbursts, the slope and normalization of
the trend of period with SW count rate is to first order the same. Using a
function of the form $P=P_0\, I_{\rm EUV}^{-\alpha}$, where $I_{\rm EUV}$ is
the 75--120~\AA \ SW count rate, the combined data is fit with parameters
$P_0=7.08\pm 0.03$~s and $\alpha = 0.09385\pm 0.00003$ if the errors on the
period are set to produce $\chi_\nu ^2= 1$, but a more realistic estimate
for the exponent is $\alpha = 0.094\pm 0.030$.

Second, there is no ``hysteresis'' in the trajectory of the 1994 outburst---it
doubles back on itself instead of ``looping,'' as do plots of the period of
optical DNOs versus optical flux (see, e.g., \cite{pat81}). This figure makes
clear that to first order the period of the EUV DNOs of SS~Cyg is determined
solely by the SW count rate. Because the SW spectrum does not change during the
observations, the bolometric correction for the EUV flux is fixed. We conclude
that the period of the EUV DNOs of SS~Cyg is determined by the EUV/soft X-ray
luminosity, and, by inference, by the mass-accretion rate onto the white dwarf.

Yet another conclusion follows from the above result that the period of the
EUV DNOs of SS~Cyg is a single-valued function of the EUV flux. Optical and
EUV light curves of SS~Cyg (\cite{mau6a}), U~Gem (\cite{lon96}), and VW~Hyi
(\cite{mau6b}) demonstrate not only the well-know result that the optical flux
leads the EUV flux by $\sim 1$ day on the rise to outburst, but that the optical
flux lags the EUV flux on the decline from outburst. With the (as-yet unproven!)
assumption that the period of the optical and EUV DNOs are equal, the looping
trajectory of the period of optical DNOs versus optical flux is understood
to be simply due to the time-dependent lead/lag of the optical flux relative
to the EUV flux, and, by extension, the mass-accretion rate onto the white
dwarf. [Quite happily, just such an effect is predicted by the accretion
disk limit cycle mechanism for dwarf nova outbursts (see, e.g., Figure 13 of
\cite{can93}).] This result, combined with the high coherence, amplitude, and
luminosity of the EUV DNOs, strongly supports the long-held belief that the
optical DNOs are produced by reprocessing of the EUV DNOs.

\section{Discussion}

Many models have been considered to explain the DNOs of high-$\Mdot $ CVs
(\cite{war95}). The low period stability of the oscillations rules out the
rotating white dwarf (the DQ~Her mechanism) as well as non-radial pulsations of
the white dwarf as the cause of the oscillations; pulsations are observed in
high inclination systems, ruling out the eclipse by the white dwarf of luminous
blobs of material in the inner disk and boundary layer; r-modes and trapped
g-modes fail because more than one mode would be excited; oscillations of the
accretion disk fail because they are not confined to a particular annulus and
hence to a particular period. Viable mechanisms are more difficult to construct.
Warner \& Livio (1996) propose that the oscillations are due to the combined
action of the differentially rotating surface layers of the white dwarf and
magnetically controlled accretion.

Despite any other indication that the white dwarf in SS~Cyg is magnetic, we
consider the requirements of a magnetospheric model for the DNOs of SS~Cyg.
Such a model has been recently applied by Finger et~al.\ (1996) to the
quasi-periodic oscillations of the hard X-ray flux of the X-ray transient
A0535+262. The Burst and Transient Source Experiment onboard the {\it Compton
Gamma-Ray Observatory\/} observed a ``giant'' outburst of A0535 during 1994
February 3 through March 20. Finger et~al.\ applied a magnetospheric model to
the pulsar in A0535 to explain the observed variations in the QPO frequency $\nu
_{\rm QPO}$ and spin-up rate of the neutron star. The rationale is a follows:
the neutron star accretes through a disk; the disk will be disrupted by the
magnetic field of the pulsar at some radius $r_0$; the observed QPOs will be
generated by some mechanism at either (a) the Keplerian frequency $2\pi
\nu_K(r_0) =(G\M* /r_0^3)^{1/2}$ at $r_0$ or (b) the beat frequency $\nu_{\rm
beat}(r_0)\equiv \nu _K(r_0) - \nu_*$ between the Keplerian frequency at $r_0$
and the spin frequency of the neutron star $\nu_*$; and torques on the
magnetosphere interior and exterior to the corotation radius $r_{\rm co}\equiv
[G\M* /(2\pi\nu_*)^2]^{1/3}$ will spin the neutron star up or down as $r_0$
varies with the mass-accretion rate. Assuming a dipole magnetic field
$B(r)=\mu/r^3$ where $\mu =B(\R* )\R* ^3$ is the dipole moment, and determining
$r_0$ by conservation of angular momentum for a  Keplerian flow (\cite{gho91}),
$$
r_0=K\, (G\M* )^{-1/7} \mu ^{4/7} \Mdot ^{-2/7},
\eqno(1)
$$
where $K$ is a constant of order unity. Hence, $\nu_K(r_0)\propto\Mdot ^{3/7}$.
With these relationships and a simple estimate for the accretion torque, Finger
et~al.\ could fit the observed variations in the QPO frequency and neutron
star spin-up rate as a function of the measured 20--100~keV flux by identifying
the QPO frequency with either $\nu _K(r_0)$ or $\nu _{\rm beat}(r_0)$.
Specifically, $\nu_{\rm QPO}\propto I_{\rm HX}^{0.43}$ over a variation of a
factor of $\approx 10$ in the observed 20--100~keV hard X-ray flux $I_{\rm
HX}$. For subsequent reference, we note here that for the parameters derived
by Finger et~al., $\mu\approx 1\times 10^{31}~\rm G~cm^3$ and, at the peak of
the outburst, $\Mdot\approx 4\times 10^{-9}~\Msun~\peryr $. With these values,
the above equation gives $r_0\approx 1\times 10^9~{\rm cm}\approx 1\times 10^3$
neutron star radii. The strength of the magnetic field on the surface of the
star is $B(\R* )\approx 1\times 10^{13}$~G and its value at $r_0$ is $B(r_0)
\approx 1\times 10^4$~G.

Whereas $\nu _{\rm QPO}\propto I_{\rm HX}^{0.43}$ for A0535, $\nu _{\rm DNO}
\propto I_{\rm EUV}^{0.094}$ for SS~Cyg. If a magnetospheric model applies to
SS~Cyg as it does to A0535, why is the variation of the DNO frequency such a
weak function of the EUV flux? Variations in the bolometric correction relating
the observed EUV flux to the EUV/soft X-ray luminosity and hence to $\Mdot $ 
are ruled out because the EUV spectrum does not change throughout the outburst.
Two alternatives exist, both of which appeal to different scalings of $r_0$
with $\Mdot $ than shown above: $r_0\propto\Mdot ^{-\alpha }$ where $\alpha
\approx 0.29$. The first alternative involves the accretion disk. Ghosh
(1996) considered four standard disk models and found $\alpha = 0.15$
for a one-temperature, optically thick, radiation pressure dominated disk;
$\alpha = 0.25$ for a one-temperature, optically thick, gas pressure dominated
disk; and larger values of $\alpha $ for more exotic two-temperature optically
thin disks. The data for A0535 are well fit by the scaling for the
one-temperature, optically thick, gas pressure dominated disk: $r_0\propto\Mdot
^{-0.25}$, hence $\nu_K(r_0)\propto \Mdot ^{0.38}$. Although a one-temperature,
optically thick, radiation pressure dominated disk gives a weaker scaling for
$\nu_K(r_0)$ with $\Mdot $, it is very unlikely that such a model describes the
disks of high-$\Mdot $ CVs.

If the structure of the inner disk of SS~Cyg does not differ significantly
from that of A0535, perhaps the structure of its magnetosphere does. The weak
dependence of the frequency of the DNOs of SS~Cyg on EUV flux suggests that
the magnetosphere of SS~Cyg is ``stiffer'' than the dipole magnetosphere
of A0535. Assuming instead of a dipole magnetic field a single star-centered 
multipole field $B(r)=m_l/r^{l+2}$ where $m_l=B(\R* )\R* ^{l+2}$ is the
multipole moment (\cite{aro93}),
$$
r_0=K^\prime \, (G\M* )^{-1/(4l+3)} m_l^{4/(4l+3)} \Mdot ^{-2/(4l+3)},
\eqno(2)
$$
hence, $\nu_K(r_0)\propto \Mdot ^{3/(4l+3)}$. To match the observed behavior
of SS~Cyg, we require $l = 7^{+4}_{-2}$. At the peak of the outburst,
the EUV/soft X-ray luminosity of SS~Cyg is $L\approx 2\times 10^{33}~\rm
erg~s^{-1}$ for $kT=20$~eV (\cite{mau95}), hence $\Mdot = 4\R* L/G\M* \approx
3\times 10^{-10} ~\Msun~\peryr $, where we have used $\M* =1.2~\Msun $ and $\R*
=3.9\times 10^8$~cm. Identifying $\nu _{\rm DNO}$ with $\nu _K(r_0)$, $r_0=
5.9\times 10^8~{\rm cm}\approx 1.5$ white dwarf radii (if instead we had used
$\M* =1.0~\Msun $, for which $\R* =5.5\times 10^8$~cm, $r_0= 5.6\times 10^8~{\rm
cm}\approx 1.0$ white dwarf radii; the magnetosphere is crushed to the surface
of the white dwarf). With these values and the above equation, the strength of
the magnetic field at $r_0$ is $B(r_0)\approx (G\M* )^{1/4} \Mdot ^{1/2}
r_0^{-5/4}\approx 5\times 10^3$~G (independent of $l$!) and its value on the
surface of the white dwarf is $B(\R* ) = B(r_0)\, (r_0/\R* )^{l+2}\approx
1\times 10^4\, (1.5)^l ~{\rm G}\sim 2^{+10}_{-1}\times 10^5$~G.

A surface magnetic field strength of 0.1--1~MG is sufficiently low to be
impossible to detect directly. In the presence of the light from the disk and
secondary, field strengths below $\approx 3$--5~MG are undetectable via circular
polarization measurements in the optical or infrared (\cite{sto92}). Direct
detection of the cyclotron harmonics is difficult because the cyclotron
fundamental is at $\lambda = 0.54\, (B/0.2~{\rm MG})$~mm. Zeeman splitting
broadens spectral lines by a measurable $\Delta\lambda/\lambda = 1.2\times
10^{-3}\, (\lambda/6563~{\rm \AA })\, (B/0.2~{\rm MG})$, but Stark broadening
is significant in the high-temperature, high-density photosphere of the white
dwarf, and the disk contributes lines with Doppler widths $\Delta\lambda/\lambda
= 8.4\times 10^{-3}\, (r/10^{10}~{\rm cm})^{-1/2}\, \sin i$. Furthermore, unlike
some other dwarf novae (e.g., U~Gem, VW~Hyi), the white dwarf in SS~Cyg is not
observed above the disk and secondary during quiescence, even in the UV. 

Of more concern is the high order required of the multipole magnetic field.
This comes about because of the weak dependence of the frequency of the DNOs of
SS~Cyg on EUV flux and, by inference, the slow variation of $r_0$ with $\Mdot
$.  We have produced this slow variation by increasing the sensitivity of
the magnetic field strength on radius: $B(r)=m_l/r^{l+2}$ with $l\sim 7$
instead of the usual expression with $l=1$. In general, the magnetic field
can be expressed as an infinite sum of multipoles: $B(r)=\Sigma _{l=1}^\infty
m_l/r^{l+2}\equiv \Sigma _{l=1}^\infty B_l(r)$. The behavior of SS~Cyg
requires that at $r_0$ the strength of the $l\sim 7$ multipole component of the
magnetic field be significantly greater than the sum of all other components:
$B_7(r_0)\gg B_l(r_0)$ for $l\ne 7$. This condition is quite restrictive, as it
requires that the corresponding surface fields $B_l(\R* )\ll (\R* /r_0)^{7-l}\,
B_7(\R* )\sim (1.5)^{l-7}\, B_7(\R* )$. It is unlikely that such a strong
inversion in the distribution of multipole components arises naturally---the
topology of the currents required to produce such a field would be complex,
to say the least.

It is much more likely that the intrinsic magnetic field of the white dwarf 
of SS~Cyg is a low-order multipole. The magnetic fields of AM Her stars are
predominantly dipolar, though often the field strengths at the two poles
are not equal, indicating that the dipole is off-center or equivalently 
that a few higher order terms enter into the expansion of the field. If the 
intrinsic magnetic field of SS~Cyg is similarly predominantly dipolar, we 
need to explain why its magnetosphere behaves so much more ``stiffly'' than 
a dipole. A clue comes from the relative sizes of the magnetospheres of
SS~Cyg and A0535. In the latter, the magnetosphere extends to $\approx 1000$
stellar radii at the peak of the outburst, while in the former it extends to 
only $\approx 1.5$ stellar radii. Suppose that the accretion disk of SS~Cyg
is sufficiently diamagnetic as to exclude a significant fraction of the 
magnetic field of the white dwarf. During the outburst, the magnetosphere
will be pinched inward by the accretion disk as the mass-accretion rate
increases and the inner edge of the disk moves inward. Trapped in the orbital
plane between the inner edge of the disk at $r_0$ and the surface of the white
dwarf at $\R* $, the magnetosphere will become significantly distorted if the
ratio $r_0/\R* $ becomes small, as it does in SS~Cyg. Expressing the distorted
field as a sum of multipoles, it is clear that higher-order terms will become
more and more important as the field is increasingly distorted. We do not
imagine that the $l\sim 7$ term ever actually dominates the multipole
expansion of the field, but that the magnetosphere comes to behave as if it did.
In the context of the simply theory behind equation (1), we require simply that
conditions conspire to produce $B(r_0)\propto r_0^{-(l+2)}$ with $l\sim 7$.
Our derived value of $B(r_0)\approx 5\times 10^3$~G should still be a
reasonable estimate, but the extrapolation of the strength of the field from
$r_0$ to the surface of the white dwarf is very uncertain. The simply scaling
gives $B(\R* )\approx B(r_0)\, (r_0/\R* )^{l+2}\approx 1\times 10^4\,
(1.5)^l~{\rm G}\sim 2^{+10}_{-1}\times 10^5$~G for $l=7^{+4}_{-2}$,
but this can be considered only an order-of-magnitude estimate.

\section{Conclusions}

If a magnetospheric model applies to SS~Cyg, it appears likely that the surface
magnetic field strength of its white dwarf is 0.1--1~MG. This range of values
is orders of magnitude lower than the field strengths of AM~Her stars ($B
\approx 10$--80~MG; \cite{cro90}; \cite{beu96}), but may overlap with the field
strengths of DQ~Her stars (very uncertain, by $B\sim 0.1$--10~MG; \cite{pat94}).
With such a field strength, it is a challenge for SS~Cyg in quiescence not to
manifest the photometric variations associated with DQ~Her stars; the limit on
such variations in the optical is 0.001 mag ($\approx 0.01\%$) near 0.1~Hz
(\cite{pat95}). In outburst, the accretion flow should be channeled down to the
footpoints of the magnetic field and produce hard and soft X-rays in the manner
of AM~Her and DQ~Her stars. However, neither the eponymous DQ~Her nor V533~Her
are hard X-ray sources (\cite{cor81}), demonstrating that magnetic accretion can
take place without the production of hard X-rays. While SS~Cyg in outburst is a
{\it known\/} hard X-ray source (\cite{jon92}; \cite{nou94}; \cite{pon95}), it
is {\it not\/} known to oscillate in hard X-rays; the limit is 6\% when soft
X-ray oscillations were observed by {\it HEAO 1\/} (\cite{swa79}). To produce
a more stringent upper limit, {\it XTE\/} observations are required. Planned
simultaneous observations of SS~Cyg in outburst with {\it XTE\/} and {\it
EUVE\/} will determine the shapes of the hard and soft X-ray light curves, the
correlations between the soft and hard X-ray fluxes, and the extent to which
the oscillations in the soft X-ray flux are manifest in the hard X-ray flux.

\acknowledgments

The {\it EUVE\/} observations of SS~Cyg could not have been accomplished
without the efforts of the members, staff, and director, J.\ Mattei, of the
American Association of Variable Star Observers; {\it EUVE\/} Deputy Project
Scientist Ron Oliversen; and the staffs of the Center for EUV Astrophysics
(CEA), the {\it EUVE\/} Science Operations Center at CEA, and the Flight
Operations Team at Goddard Space Flight Center. The author greatly benefited
by conversations with P.~Ghosh, J.~Patterson, R.~Robinson, and B.~Warner.
M.~Finger, P.~Ghosh, M.~Livio, and B.~Warner generously provided copies of
material prior to publication. The manuscript benefited from the comments
of an anonymous referee. This work was performed under the auspices of the
U.S.\ Department of Energy by Lawrence Livermore National Laboratory under
contract No.~W-7405-Eng-48.

\clearpage    


\clearpage    


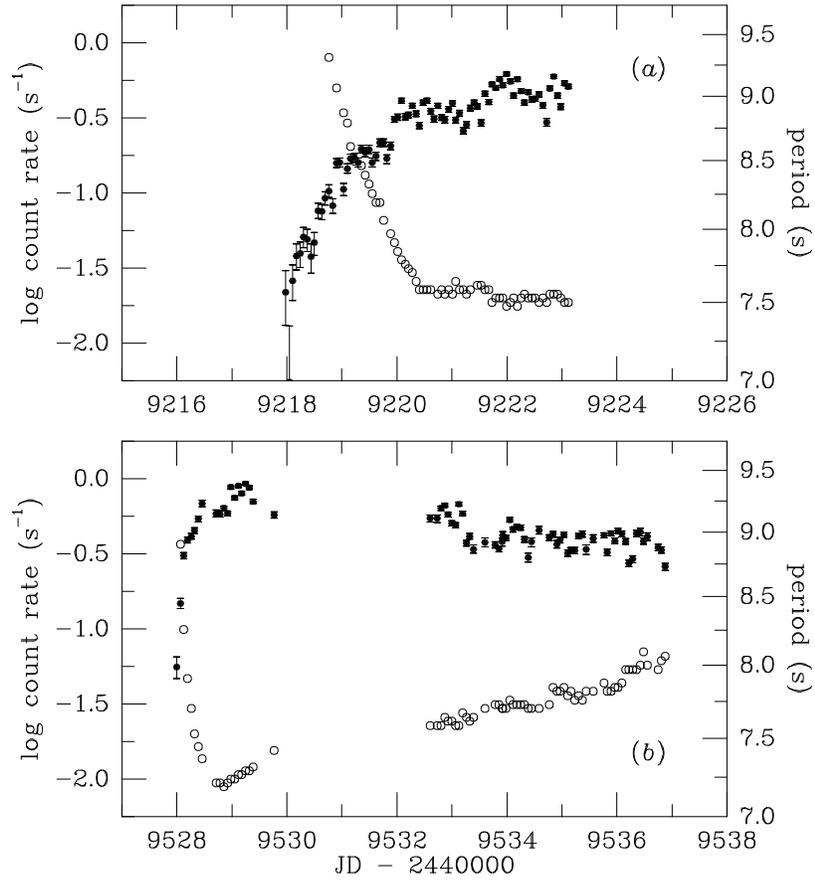
\begin{figure}
\caption{Log of the 75--120~\AA \ SW count rate ({\it filled circles with
error bars\/}) and log of the period ({\it open circles\/}) as a function of
time for the ({\it a\/}) 1993 August and ({\it b\/}) 1994 June/July outbursts
of SS~Cyg.}
\end{figure}

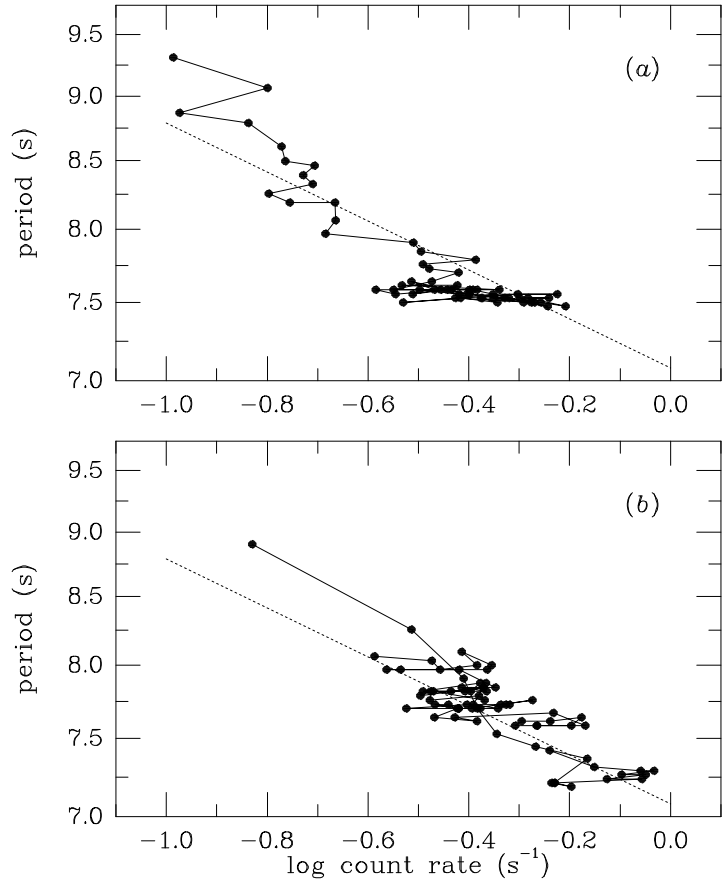
\begin{figure}
\caption{Log of the period as a function of log of the 75--120~\AA \ SW
count rate for the ({\it a\/}) 1993 August and ({\it b\/}) 1994 June/July
outbursts of SS~Cyg. The dotted line is an unweighted fit to the combined
data: $P = 7.08\, I_{\rm EUV}^{-0.094}$~s.}
\end{figure}

\end{document}